\documentclass[english,aps,a4paper,preprint,showpacs,amsfonts,amsmath,amssymb,nofootinbib]{revtex4}
\usepackage[english]{babel}
\usepackage[dvips]{graphicx}

\def\Journal#1#2#3#4#5{{#1}, {\it #2} \textbf{#3}, #4 (#5)}
\def\Book#1#2#3#4#5{{#1}, {\it #2} (#3, #4, #5)}
\def\Sref#1{Section~\ref{#1}}

\newcommand{\dd}{\mbox{d}}
\newcommand{\rr}{R}
\newcommand{\CE}{\mathcal E}
\newcommand{\EE}{\mathbb{E}}
\newcommand{\RR}{\mathbb{R}}
\newcommand{\tto}{t_0}

\begin{document}

\title{Predator-Prey Model for Stock Market Fluctuations}
\author{Miquel Montero}
\email[E-mail: ]{miquel.montero@ub.edu}
\affiliation{Departament de F\'{\i}sica Fonamental, Universitat de
Barcelona,\\  Diagonal 647, E-08028 Barcelona, Spain}
\date{\today}

\begin{abstract}
We present a dynamical model for the price evolution of financial assets. The model is based in a two level structure. In the first stage one finds an agent-based model that describes the present state of the investors' beliefs, perspectives or strategies. The  dynamics is inspired by a model for describing predator-prey population evolution: agents change their mind through self- or mutual interaction, and the decision is adopted on a random basis, with no direct influence of the price itself. One of the most appealing properties of such a system is the presence of large oscillations in the number of agents sharing the same perspective, what may be linked with the existence of bullish and bearish periods in financial markets. In the second stage one has the pricing mechanism, which will be driven by the relative population in the different investors' groups. The price equation will depend on the specific nature of the species, and thus it may change from one market to the other: we will firstly present a simple model of excess demand, and subsequently consider a more elaborate liquidity model. The outcomes of both models are analysed and compared.
\end{abstract}
\pacs{89.65.Gh, 87.23.Ge, 02.50.Ey, 05.40.Jc, 05.10.Gg}
\maketitle
\section{Introduction}
Financial models based on interacting agents possess a large tradition in the economic literature~\cite{H06} |one of the first references in which the evolution of a market is related with the activity of individual investors dates back to 1974
|, but they have gained relevance in the physics literature in relatively recent years~\cite{MM06,SZSL07}. The complete list of such models is so extensive and their properties so diverse that we can merely sketch here the recurrent traits shared by most of the models, and address the reader to the references cited in~\cite{H06,MM06,SZSL07}. 

The pioneering work of 1974~\cite{Z74} already contains one of the more ubiquitous ingredients in the subsequent agent-based models: heterogeneity. Agents are assumed to be heterogeneous to some extend, and therefore they can be aggregated into one out of a finite set of categories. Since the minimum number of different categories is two, and simplicity is often a plus, investors are usually arranged into two (competing) groups. The terms used to name them and their defining properties are uneven across the literature |Chartists and Fundamentalists~\cite{Z74}, Trend-followers and Contrarians~\cite{MGTM04},  Speculators and Producers~\cite{Z99}, Imitators and Optimizers~\cite{C80}| but the ideas lying behind are similar, and can be well represented by Chartists and Fundamentalists.
Chartists are (sometimes adaptative) agents whose inversion strategy is based in the belief that {\it past\/} information may contain clues about the {\it future\/} evolution of the security, and therefore that they can infer future prices. Fundamentalists are in essence agents that think that they can deduce the {\it present\/} value of a firm on the basis of the information {\it presently\/} available, such as dividend payments or earning rates. Fundamentalists operate in a rather predictable way since they expect that the market tends to correct any observed deviation between fundamental and market prices: they sell overpriced securities and buy underpriced ones. The picture is not so simple for Chartist-like investors since at the end they deploy rule-of-thumb strategies, like the Moving Average Convergence Divergence (MACD) indicator or the Relative Strength Index (RSI), two technical analysis tools broadly used by actual financial practitioners. 
Therefore, the list of available strategies in agent-based models may be so large that, in the most extreme situation, strategies may differ for any pair of investors in the market, as in some instances of the Minority Game market model~\cite{MM06,CZ97}.
In fact, any single agent may combine technical trading rules with fundamental ones, or decide among them, what makes evolve the profile of the investors. There is no doubt that this diversity adds more heterogeneity into the model.

Another general trait of present models is that agents are boundedly rational~\cite{S79}: they  decide their actions in the next time step on the basis of a limited and possibly incomplete amount of information. They ignore the beliefs of the rest of investors and usually cannot evaluate the consequences of their own decisions. Under these circumstances selfish agents try to maximize a pay-off or utility function, a measure of their individual success. 

The final ingredient is the pricing mechanism. The usual paradigm when agent activity does not explicitly settle the price of the asset is to define a differential equation or a finite difference equation that relates the price evolution with the relevant (global) variables of the model. Since these variables are affected by the mutual interaction of the investors in a complex way, two complementary approaches are generally considered: the behaviour of the system is computer simulated and/or the complexity is reduced by considering that the number of agents approaches to infinity, the thermodynamic limit. 

As we will shortly show, some of the previous ingredients either are not present in this agent-based model, or have been introduced with a different philosophy. The model was inspired by a previous article on population dynamics~\cite{MKN05}, where authors 
reported presence of large oscillations in the species densities due to a finite-size stochastic effect.
We export this idea into the financial language with the confection of a model that describes general aspects of investors' dynamics. 
The behaviour of the asset price is first obtained after assuming a simple model of excess demand, and subsequently, we apply the same approach to the modelling of the interaction between limit and market orders in a stock market. The overall result exhibits similarities with prevailing agent-based financial models~\cite{BG80,K93,LM99,CB00,CCMZ01}.

The paper is structured in three main Sections. \Sref{AM} deals with the agent model strictly speaking: First we define who are the agents and the three different states in which they can be found at every moment, the mechanisms that govern the changes from one state to the other, and the transition rates between states. Then we derive a master equation that characterises the time evolution of the system, and analyse the stationary solutions of this equation in the thermodynamic limit. Finally we find the second-order corrections and show their relevance in finite-size models. In \Sref{MM} we establish a first connection between the agent model and market price changes: we simulate the time evolution of the system under representative market conditions, analyse the most relevant traits, and compare them with well-known empirical properties of actual financial time series. In \Sref{Ext} we propose a second identification for the species categories (liquidity providers and liquidity takers), and a different price formation procedure is considered. The outcome presents new properties that are still consistent with what one may find in practice. This reinforces the potentials of the model.
Conclusions are drawn in \Sref{Conclusions}, and some technical derivations concerning 
correlation functions are left to an Appendix.

\section{Agent dynamics}
\label{AM}
As we have just stated, this Section deals exclusively with the intrinsic features that the agent interplay generates. For this aim, as we will see, we do not need a detailed description of the agents' inner properties. The most important point to be made here is about the motivation and plausibility of the agent-based approach introduced. 

Along this article we will assume that any trader that may ever operate in our financial market can be accommodated in two great, well-defined and excluding groups. The first and most populated group of investors will constitute what is usually termed as {\it noise traders\/}~\cite{CCMZ01,CMZ00}. We will assume that each one of these traders acts in a purely random fashion, independently of the rest of agents in the market. We are not considering this kind of traders as individuals, they merely act as a some sort of thermal bath or noise source, what reinforces the foundations of the stochastic character of the dynamics to be introduced. 

The second group of traders is the set of those which we will call {\it qualified investors\/}, but the term informed traders~\cite{H07,BDFG09} would be fit for them as well. As we are showing below, we will connect the price evolution with the collective state of these players, so we will consider that this group embraces the main actors with bigger influence in financial terms: mutual funds, investment banks, or corporations in general. The total amount of such participants in a real market is much more moderate~\cite{LMVM08}, what makes sense to consider a finite-size agent model to describe them. We will assume that the instantaneous state of these investors will fluctuate within three categories as a consequence of the interaction with the rest of agents. 

Thus, in essence, our agent model will describe how populated are these three states as a function of time. The detailed way in which the states ought to be defined will depend on the specific financial market under analysis. Different markets may be sensitive to different aspects of the agents' activity. We will delay the definitive profiling of the categories until the market dynamics is considered, in~\Sref{MM} and~\Sref{Ext}, and centre our attention now in the species self-interaction, as announced. 

\subsection{The species self-interaction}
\label{AMA}
Let us consider then a finite set of $N$ fully connected interacting agents who, at every instant of time $t$, may be found in one out of three possible states that we will label by the letters $A$, $B$ and $E$. In a very general sense, that must be further refined from case to case, we will assume that an abundance of agents in state $A$, $N_A(t)$, is related with a bear market scenario,  that the increasing of population in the $B$ side, $N_B(t)$, leads to a bull market scenario, whereas the market is not sensitive to changes in the number of agents of type $E$, $N_E(t)$, beyond the fact that $N=N_A(t)+N_B(t)+N_E(t)$ is fixed. We can think for example that these three sets categorise those agents that are either willing to sell, to buy, or that keep a veiled attitude, but there are several alternative identifications for groups $A$ and $B$ |liquidity providers and liquidity takers, or pessimistic and optimistic investors| that would equally help in visualising these two groups. In any case, within our formulation $E$ will always represent a neutral position. 

Another ingredient of the model is the fact that the mechanism which allows agents to change their minds and move from some state to another is based on self- and mutual interactions. Note specially that decisions are not affected by the previous history (there is no necessity in selling after buying, for instance) what confers the process with the Markov property.

In a population dynamics language, one has two species living in a finite world: As we will see afterwards with the description of the {\it rules of engagement\/}, within this model, the $A$'s will play the role of preys and the $B$'s will be the predators. The $E$'s, those agents without a definite or explicit intention, act as empty space.

The basic unitary interaction in population problems is the death process, $A \stackrel{p}{\rightarrow} E$ and $B \stackrel{q}{\rightarrow}E$. Each one of these two processes, and the same applies for the rest of interactions, may encompass the aggregate effect of 
disparate contributors.~\footnote{Note however that our approach for the agent dynamics is mostly phenomenological. The available interactions were selected with the aim of capturing general {\it macroscopic\/} features. Therefore the proposed and future interpretations of the processes in terms of {\it microscopic\/} events fulfil mainly illustrative purposes.} So, within an offer/demand framework we ought to include here all the transactions between the agents and the noise traders, perhaps the most common transaction instance in actual stock markets, but one can also take into account the possibility that qualified investors may decide to decline in their previous intentions by their own. In this scheme $p$ and $q$ represent the probability per unit of time that a given active agent when observed separately passes into inactivity. The same kind of notation is used in the description of the remaining transitions.

Yet another typical unitary interaction in population models in the spontaneous birth of preys, $E \rightarrow A$, but this is not considered here. All birth processes are due to those binary interactions that also occur in the system. At this point, it can be useful from a practical point of view to establish the probability $\nu$ of considering rather a two-component transition than a single-component one. 

The first two-component interaction we will consider is the following: $AB \stackrel{a}{\rightarrow} EE$. It may be specifically explained in terms of transactions among the qualified investors, but in a general sense, it conveys a form of agreement between two active investors in such a way that none of them convinces the other. This annihilation process is not usually considered in population models: it represents some sort of deadly defense in which predator and prey die after fighting. The usual result in predator-prey models after $AB$ interactions is predation: $AB \stackrel{b}{\rightarrow} BB$. In our case, this counts for the possibility that an active investor may perform a change in the evaluation of the market scenario (from bear to bull) 
due to predominance of $B$'s. This may eventually lead to a market bubble. Once again it may become useful to consider that some 
fraction $\lambda$ of $AB$ interactions conduces to annihilation, whereas $1-\lambda$ of them ends in predation.

The third binary interaction is $AE \stackrel{c}{\rightarrow} AA$, in which an agent that was not interested in operating in the market comes into activity in the $A$ side. Under financial optics this imitative behaviour can lead to market panic and ultimately to a crash. Here lies our birth mechanism for preys that also incorporates in the model population pressure against unbounded prey growth.

The complete state of the agent system at given instant $t$ is fully settled by the number of investors belonging to species $A$ and $B$, $N_A(t)$ and $N_B(t)$ respectively. Since these numbers will be stochastic magnitudes, we are interested in obtaining an expression for $P(n,m,t)$, the probability of having $n$ $A$'s and $m$ $B$'s at time $t$:
\begin{equation*}
P(n,m,t)=\Pr\{N_A(t)=n, N_B(t)=m\}. 
\end{equation*}
To this end we will consider the following five {\it transition rates\/}, the transition probabilities (per unit of time) between macroscopic states, based upon the above interactions:
\begin{eqnarray*}
T(n-1,m-1|n,m)&=&2\nu\lambda a \frac{n}{N} \frac{m}{N-1}, \\
T(n-1,m|n,m)&=&(1-\nu) p \frac{n}{N}, \\
T(n-1,m+1|n,m)&=&2\nu(1-\lambda) b \frac{n}{N} \frac{m}{N-1}, \\
T(n,m-1|n,m)&=&(1-\nu) q \frac{n}{N}, \\
T(n+1,m|n,m)&=&2 \nu c \frac{n}{N} \frac{N-n-m}{N-1}.
\end{eqnarray*}
Note that there are also three ``single-stepped'' feasible transitions on each variable which are not listed here and, therefore, that  are forbidden: $T(n,m+1|n,m)$, $T(n+1,m+1|n,m)$ and $T(n+1,m-1|n,m)$. 

Summing up, in essence we are assuming that:
\begin{enumerate}
\item states $A$ and $B$ can spontaneously decay into inactivity,
\item there is a basic non-trivial interaction (commonly the trade) that is not sensitive to the interchange of $A$ and $B$ roles,
\item $B$'s can convince $A$'s only, and
\item $A$'s can convince $E$'s only,
\end{enumerate}
where the asymmetry in the two last items 
expresses the fact that bubbles and crashes in actual stock markets are different in shape~\cite{BC98,LM00}.  Since all agents are identical, 
the heterogeneity of our model relies on this asymmetry in the interactions.

\subsection{The master equation}
\label{ME}
\begin{figure}
{
\includegraphics[height=0.80\columnwidth,keepaspectratio=true,angle=-90]{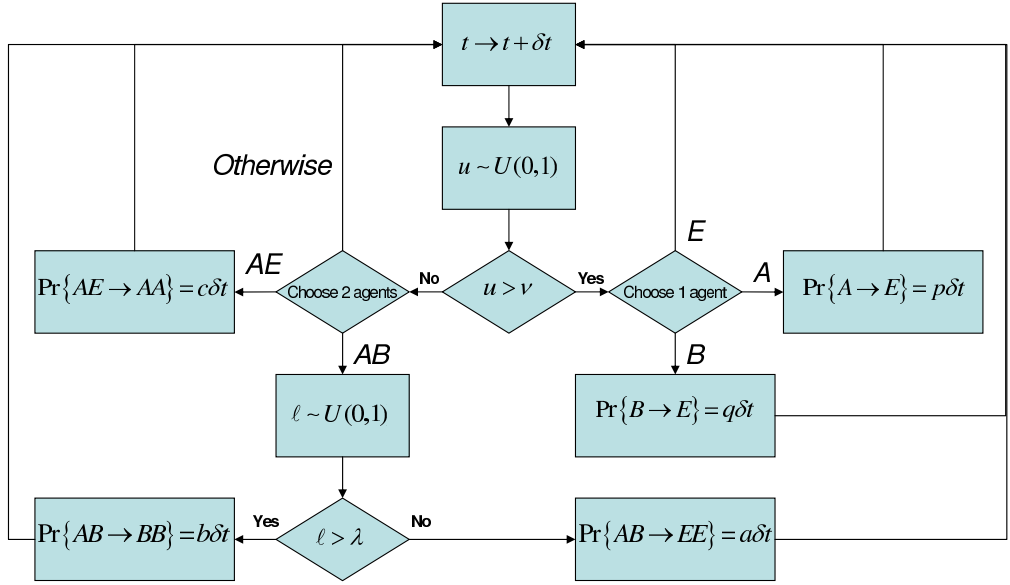}
}
\caption{(Color online) Flux diagram for a discrete-time updating procedure of the state of the agents.}
\label{Update_Fig}
\end{figure}

The Markov character of the model makes superfluous considering more sophisticated transition rates in the elaboration of the master equation (ME), 
the equation that defines the time evolution of $P(n,m,t)$:
\begin{eqnarray*}
\frac{\dd P(n,m,t)}{\dd t}&=& (\alpha_{AA}-\gamma_A) (\CE_x^{+1}-1)[n P(n,m,t)] \\
&+&\gamma_B(\CE_y^{+1}-1)[m P(n,m,t)]\\
&+& \frac{\alpha_{AB}-\beta_{AB}-\alpha_{AA}}{2} (\CE_x^{+1}\CE_y^{+1}-1)\left[n \frac{m}{N-1} P(n,m,t)\right]\\
&+&\frac{\alpha_{AB}+\beta_{AB}-\alpha_{AA}}{2} (\CE_x^{+1}\CE_y^{-1}-1)\left[n \frac{m}{N-1} P(n,m,t)\right]\\
&+& \alpha_{AA} (\CE_x^{-1}-1)\left[n \frac{N-n-m}{N-1} P(n,m,t)\right].
\end{eqnarray*}
Here we have introduced the following increment/decrement operators
\begin{eqnarray*}
 {\cal E}_x^{ \pm 1} f(n,m,t) \equiv f(n \pm 1,m,t), \\ 
 {\cal E}_y^{ \pm 1} f(n,m,t) \equiv f(n,m \pm 1,t), \\ 
\end{eqnarray*}
and five new parameters
\begin{eqnarray}
\gamma_A &=& \frac{2\nu c-(1-\nu) p}{N},
\label{gamma_A}\\
\gamma_B&=& \frac{(1-\nu) q}{N},
\label{gamma_B}\\
\alpha_{AA}&=&\frac{2 \nu c}{N},
\label{alpha_AA}\\
\alpha_{AB}&=&2 \nu \frac{\lambda a+(1-\lambda)b +c}{N}, 
\label{alpha_AB}\\
\beta_{AB}&=&2 \nu \frac{(1-\lambda)b-\lambda a}{N},
\label{beta_AB}
\end{eqnarray}
which encode all the relevant information of the model parameterisation. Let us stress that $\lambda$ and $\nu$ were defined in order to clarify how the update mechanism can be approximately implemented, see Fig.~\ref{Update_Fig}, but they do not introduce further degrees of freedom to the problem since they would disappear after a constant redefinition. This is the case if one uses Gillespie's exact algorithm~\cite{G76} in the simulation of the system, as we have done.

The relevance of the new parameters becomes noticeable soon afterwards. Suffice it to say for the moment that we will proceed as they were independent of the size of the system in what follows, because we will consider the expansion of the ME in terms of powers of $N$. 
To this end let us define $R_{A,B}(t)$, 
\begin{equation*}
R_{A,B}(t)\equiv \lim_{N\rightarrow \infty}\EE[N_{A,B}(t)]/N.
\end{equation*}
and introduce two new stochastic processes, $X(t)$ and $Y(t)$, in such a way that
\begin{eqnarray*}
N_A(t) &=& N\ R_A(t)  + \sqrt N X(t),\\
N_B(t) &=& N\ R_B(t)  + \sqrt N Y(t),
\end{eqnarray*}
hold. $X(t)$ and $Y(t)$ are thus responsible for the {\it fluctuations\/} of $N_A(t)$ and $N_B(t)$ around their mean values. It is expected that the strength of those fluctuations will diminish as the system reaches the thermodynamic limit, that is, when $N \gg 1$. 
Note that this approach implies that, for any two given values of the species population, $n$ and $m$, we will have that 
\begin{eqnarray*}
n &=& N\ R_A(t)  + \sqrt N x,\\
m &=& N\ R_B(t)  + \sqrt N y,
\end{eqnarray*}
where $x$ and $y$ |as well as $R_A(t)$ and $R_B(t)$| are real magnitudes in spite that $n$ and $m$ were integers. In such a situation increment/decrement operators become partial differential operators~\cite{VK92},
\begin{eqnarray*}
 {\cal E}_x^{ \pm 1}  = 1 \pm \frac{1}{{\sqrt N }}\partial _x  + \frac{1}{{2N}}\partial _{xx}^2  + {\cal O}(N^{ - 3/2} ),\\ 
 {\cal E}_y^{ \pm 1}  = 1 \pm \frac{1}{{\sqrt N }}\partial _y  + \frac{1}{{2N}}\partial _{yy}^2  + {\cal O}(N^{ - 3/2} ).
\end{eqnarray*}
Finally note that $P(n,m,t)$ must be replaced by $\Pi(x,y,t)$,
\begin{equation*}
\Pi(x,y,t)\dd x \dd y \equiv \Pr\{x<X(t)\leqslant x+\dd x,y<Y(t)\leqslant y+\dd y\},
\end{equation*}
through
\begin{equation*}
P(n,m,t) 
=\frac{1}{N}\Pi\left(\frac{n-N \rr_A}{\sqrt{N}},\frac{m-N \rr_B}{\sqrt{N}},t\right)\dd x\dd y,
\end{equation*}
what affects the time derivative term in the ME in the following way:
\begin{equation*}
\frac{\dd P}{\dd t}
=-\left[\frac{1}{\sqrt{N}}\frac{\dd \rr_A}{\dd t}\partial_x \Pi +\frac{1}{\sqrt{N}}\frac{\dd \rr_B}{\dd t}\partial_y \Pi-\frac{1}{N}\partial_t \Pi\right]\dd x \dd y.
\end{equation*}

\subsection{First-order stationary solutions}
\label{AMB}
The first-order approximation of the ME collects terms of order $N^{ - 1/2}$, ignores those of ${\cal O}(N^{ -1})$, and leads to a set of coupled Volterra equations for $\rr_A(t)$ and $\rr_B(t)$,
\begin{eqnarray}
\frac{\dd \rr_A}{\dd t} &=& \left[\gamma_A - \alpha_{AA} \rr_A -\alpha_{AB} \rr_B \right]\rr_A,  \label{rr_A}\\
\frac{\dd \rr_B}{\dd t} &=& \left[\beta_{AB} \rr_A-\gamma_B\right]\rr_B. 
\label{rr_B}
\end{eqnarray}
Let us analyse the factors appearing in these equations. $\gamma_A$ as defined in 
Eq.~(\ref{gamma_A}) represents a trade-off between a positive term that comes from the {\it imitation\/} influence and a negative term that measures the death rate of preys. If positive, it would correspond to a effective birth rate of preys in Eq.~(\ref{rr_A}). 
Recall however that in this system preys suffer of population pressure instigated by the imitation interaction that constrains the preys' growth, see the definition of $\alpha_{AA}$ in~(\ref{alpha_AA}). The term with the $\alpha_{AB}$ factor counts for the reduction in the prey number due to all binary operations, not only predation, Eq.~(\ref{alpha_AB}). The $\beta_{AB}$ term appearing in Eq.~(\ref{rr_B}) is a consequence of the imbalance between predation and annihilation alternatives, as it can be observed in~(\ref{beta_AB}), whereas $\gamma_B$ measures exclusively the death rate of predators, expression~(\ref{gamma_B}). Summing up, there are two parameters, $\gamma_A$ and $\beta_{AB}$, with no definite sign, whereas
$\gamma_B$, $\alpha_{AA}$ and $\alpha_{AB}$ are positive constants {\it ab initio\/}. 

Eqs.~(\ref{rr_A}) and~(\ref{rr_B}) present three stationary solutions for which
\begin{equation*}
\frac{\dd \rr_A}{\dd t}=\frac{\dd \rr_B}{\dd t}=0.
\end{equation*}
The first solution is the trivial one, $\rr_A=\rr_B=0$. It represents the death of the market due to a complete lack of activity. This is a feasible scenario that threatens any real market. For instance, investors may loose interest in any given 
commodity that becomes useless or exhausted. The stability analysis of this fixed point determines that it will be a saddle point if $\gamma_A>0$, otherwise it would turn stable.
The analysis of the 
second stationary solution,
$\rr_A=\gamma_A/\alpha_{AA}\equiv M/N<1$ |note that $\gamma_A<\alpha_{AA}$ by construction,  {\it cf.\/} expressions~(\ref{gamma_A}) and~(\ref{alpha_AA})|, and $\rr_B=0$, leads to the constraint 
\begin{equation}
0<\frac{\gamma_B}{\beta_{AB}}<\frac{M}{N}, 
\label{beta_AB_const}
\end{equation}
if one wants to avoid conferring stability to this fixed point as well.~\footnote{Note how this situation may represent the worst market crash imaginable, where every active investor is in the bear side, what would conclude in a liquidity crisis like the present  one. In this sense one may consider that $M$ is related to the finiteness of the total amount of shares.} 
In conclusion, all the parameters defined in (\ref{gamma_A})-(\ref{beta_AB}) must be positive-definite. 

We must point out that the presence of those unstable equilibrium solutions is not a flaw but a merit of the model, as is the fact that the remaining stationary solution
\begin{eqnarray*}
\rr_A=\rr_A^\circ &\equiv& \frac{\gamma_B}{\beta_{AB}},\\
\rr_B=\rr_B^\circ &\equiv& \frac{\gamma_A \beta_{AB}-\gamma_B \alpha_{AA}}{\alpha_{AB} \beta_{AB}}, 
\end{eqnarray*}
is always present, accessible and corresponds to a stable fixed point. 

Regarding the occurrence of the fixed point, it is evident that $\rr_A^\circ>0$, and Eq.~(\ref{beta_AB_const}) leads to $\rr_A^\circ<M/N<1$.
The same equation determines that $\rr_B^\circ>0$. Also $\rr_B^\circ<1$, as it can be proven as follows:
\begin{equation*}
\rr_B^\circ = 1- \frac{(\alpha_{AB}-\gamma_A) \beta_{AB}+\gamma_B \alpha_{AA}}{\alpha_{AB} \beta_{AB}}<1, 
\end{equation*}
because trivially $\alpha_{AB}>\gamma_A$, {\it cf.\/} Eqs.~(\ref{gamma_A}) and~(\ref{alpha_AB}). We can also show that $\rr_A^\circ+\rr_B^\circ<1$,
\begin{eqnarray*}
\rr_A^\circ+\rr_B^\circ &=& 1- \frac{(\alpha_{AB}-\gamma_A) \beta_{AB}+ (\alpha_{AA}-\alpha_{AB})\gamma_B}{\alpha_{AB} \beta_{AB}}\\
&<&1- \frac{\gamma_B}{\gamma_A}\frac{(\alpha_{AB}-\gamma_A) \alpha_{AA}+ (\alpha_{AA}-\alpha_{AB})\gamma_A}{\alpha_{AB} \beta_{AB}}\\
&=&1- \frac{\gamma_B(\alpha_{AA}-\gamma_A)}{\gamma_A \beta_{AB}}<1, 
\end{eqnarray*}
because $\alpha_{AA}>\gamma_A$ as we have just pointed out above. 

The analysis of the stability of this fixed point leads to the conclusion that the point is stable, and that the transient term will exhibit oscillations when $\omega_0\in \RR^+$,
\begin{equation}
\omega_0=\sqrt{\alpha_{AB} \beta_{AB} \rr_A^\circ \rr_B^\circ-\frac{1}{4}\left(\alpha_{AA} \rr_A^\circ\right)^2},
\label{omega0}
\end{equation}
which is true whenever
\begin{equation*}
\frac{\alpha_{AA}}{\beta_{AB}}<2 \sqrt{1+\frac{\gamma_A}{\gamma_B}}-2. 
\end{equation*}
When the system shows transient oscillations, there is a single characteristic time scale for the decay rate, 
\begin{equation}
\tau_0=\frac{2}{\alpha_{AA} \rr_A^\circ},
\label{tau_0}
\end{equation}
and for $t\gg \tau_0$ the system would reach the stable solution. This assertion is no longer true when a second decay rate appears, since
\begin{equation}
T_0^{-2}=\tau_0^{-2}-\alpha_{AB} \beta_{AB} \rr_A^\circ \rr_B^\circ<\tau_0^{-2}. 
\label{T0}
\end{equation}
In such a situation the steady state is reached when $t^{-1} \ll \tau_0^{-1}-T_0^{-1}$. Therefore we may define $\tto$, $\tto^{-1} \equiv \tau_0^{-1}-\Re[T_0^{-1}]$, and the steady state is always achieved for $t\gg \tto$. 

\begin{figure}
{
\includegraphics[width=0.75\columnwidth,keepaspectratio=true]{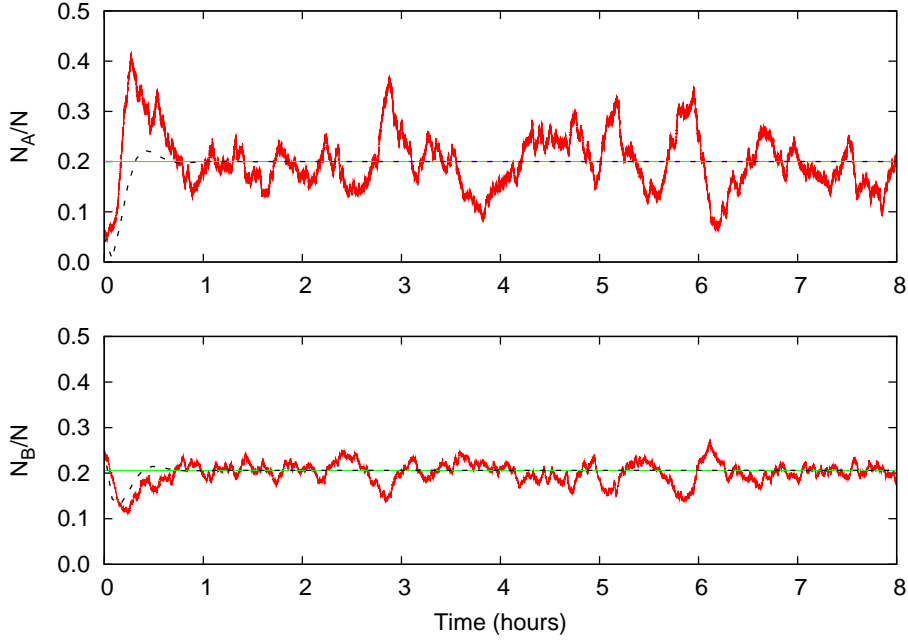}
}
\caption{(Color online) Time evolution of preys and predators densities (solid red line) for an exact realisation of our interacting agent model with $N=1000$. The dashed black line depicts the first-order approach to the problem, whereas in green is shown the stationary solution. 
We can see how fluctuations in both populations are larger than expected.}
\label{Spec_Fig}
\end{figure}

After the transient regime, and whenever $N$ is finite, we will await that the time evolution of prey and predator densities, $N_A(t)/N$ and $N_B(t)/N$, makes them attain their limit values $\rr_A^\circ$ and $\rr_B^\circ$, and exhibit some fluctuating activity afterwards. Since the characteristic size of the fluctuations is of order $N^{-1/2}$, 
a naive analysis could lead to the conclusion that if we have, let us say, $1000$ interacting agents the error in neglecting the remaining terms in the ME should be around $3.2\%$. In Fig.~\ref{Spec_Fig} we can find the outcome of a realisation of the model with $N=1000$ |the complete set of parameter specifications is listed below in \Sref{MM}.  
The example shows that in such a system fluctuations may be larger than expected, and further corrections to the first-order equations must be taken into account~\cite{MKN05,CM03}. 

A final word on the $N$ dependency of the above expressions before exploring the forthcoming terms in the ME. The order-by-order analysis under progress relies on the fact that the parameters defined in Eqs.~(\ref{gamma_A})-(\ref{beta_AB}) are independent of $N$. Note however that the expressions for $\rr_{A}^\circ$, $\rr_{B}^\circ$ and $M/N$ are insensible to this need. It only affects those constants where time is involved, like $\tau_0$ or $\omega_0$.

\subsection{Beyond the first-order equations}
\label{AMC}
When one gathers the terms of order $N^{-1}$ in the ME expansion, a  Fokker-Planck equation for $\Pi(x,y,t)$ emerges:
\begin{eqnarray*}
\partial_t \Pi&=&\left[-\gamma_A  + 2 \alpha_{AA} \rr_A+\alpha_{AB} \rr_B\right] \partial_x (x \Pi) \\
&+& \alpha_{AB} \rr_A \partial_x(y \Pi) -\beta_{AB} \rr_B \partial_y(x \Pi)+\left[\gamma_B-\beta_{AB} \rr_A\right] \partial_y(y \Pi)\\
&+&\frac{\rr_A}{2}\left[-\gamma_A + \alpha_{AA} (2- \rr_A - 2\rr_B) + \alpha_{AB} \rr_B \right] \partial^2_{xx} \Pi \\
&+& \frac{\rr_B}{2}\left[\gamma_B + (\alpha_{AB}-\alpha_{AA}) \rr_A   \right] \partial^2_{yy} \Pi - \beta_{AB} \rr_A \rr_B \partial^2_{xy} \Pi.
\end{eqnarray*}
If we concentrate our analysis of the previous equation for times large enough to let $\rr_A(t)$ and $\rr_B(t)$ reach their steady state values, $\rr_A^\circ$ and $\rr_B^\circ$, the expression simplifies considerably:
\begin{eqnarray*}
\partial_t \Pi&=&\mu_{xx} \partial_x (x \Pi) + \mu_{xy} \partial_x(y \Pi) -\mu_{yx} \partial_y(x \Pi)\\&+&\frac{1}{2} \sigma^2_x \partial^2_{xx} \Pi + \frac{1}{2}\sigma^2_y \partial^2_{yy} \Pi - \rho \sigma_x \sigma_y \partial^2_{xy} \Pi,
\end{eqnarray*}
with 
\begin{eqnarray*}
\mu_{xx}&\equiv&\alpha_{AA} \rr_A^\circ=\frac{2}{\tau_0},\\
\mu_{xy}&\equiv&\alpha_{AB} \rr_A^\circ,\\ 
\mu_{yx}&\equiv&\beta_{AB} \rr_B^\circ,\\
\sigma^2_x&\equiv&2\alpha_{AA}\rr_A^\circ\left(1- \rr_A^\circ - \rr_B^\circ \right),\\
\sigma^2_y&\equiv&\rr_A^\circ \rr_B^\circ\left(\beta_{AB} + \alpha_{AB}-\alpha_{AA} \right),\\
\rho&\equiv&\beta_{AB}\frac {\rr_A^\circ \rr_B^\circ}{\sigma_x \sigma_y}, 
\end{eqnarray*}
positive-definite quantities.~\footnote{Note in particular that $\beta_{AB} + \alpha_{AB}-\alpha_{AA}=4 \nu (1-\lambda)b/N>0$.} Therefore we have a linear multivariate Fokker-Planck equation for the joint probability density of $X(t)$ and $Y(t)$, whose solution can be systematically obtained after some (or maybe plenty of) algebra~\cite{VK92}. An alternative approach is based in the following set of coupled (It\^o) stochastic differential equations:
\begin{eqnarray}
\dd X &=&-\mu_{xx} X\dd t - \mu_{xy} Y \dd t+ \sigma_x \dd W_1, \label{dX}\\
\dd Y &=& \mu_{yx} X \dd t-\rho \sigma_y \dd W_1 + \sigma_y \sqrt{1-\rho^2}\dd W_2, \label{dY}
\end{eqnarray}
where $W_1$ and $W_2$ are two independent Wiener processes.

\subsection{The magnifying effect}
In order to explore the reason for the abnormal magnitude of fluctuations we should compare $X(t)$ and $Y(t)$ with $\rr_A^\circ$ and $\rr_B^\circ$ respectively. A quick analysis reveals that mean values are not useful in this venture because $\lim_{t\rightarrow \infty} \EE[X(t)]=\lim_{t\rightarrow \infty} \EE[Y(t)]=0$ |remember that (\ref{dX}) and (\ref{dY}) are valid for $t\gg\tto$. We concentrate in variances and co-variances instead. In Appendix~\ref{AA} we can find how the stationary values of $\EE[X^2(t)]$, $\EE[Y^2(t)]$, and $\EE[X(t)Y(t)]$ follow:
\begin{eqnarray*}
C_{xx}(0)&=&\lim_{t\rightarrow \infty} \EE[X^2(t)]=\frac{\mu_{yx} \sigma^2_x+ \mu_{xy} \sigma^2_y}{2 \mu_{xx}\mu_{yx}}, \\
C_{yy}(0)&=&\lim_{t\rightarrow \infty} \EE[Y^2(t)]=
\frac{\mu_{yx}^2 \sigma_x^2 +\left(\mu_{xx}^2 +\mu_{xy} \mu_{yx}\right)\sigma^2_y-2\rho \mu_{xx}\mu_{yx}\sigma_x \sigma_y}{2 \mu_{xx}\mu_{xy}\mu_{yx}}, \\
C_{xy}(0)&=&\lim_{t\rightarrow \infty} \EE[X(t)Y(t)]=-\frac{\sigma^2_y}{2 \mu_{yx}},
\end{eqnarray*}
and fluctuations will be typically large if these quantities are much more bigger than $(\rr_A^\circ)^2$, $(\rr_B^\circ)^2$ and $\rr_A^\circ \rr_B^\circ$, respectively. If we define the magnifying factors $\Omega_{xx}$, $\Omega_{yy}$ and $\Omega_{xy}$ as the corresponding quotient of these magnitudes, {\it e.g.\/} 
\begin{eqnarray*}
\Omega_{xy}\equiv\lim_{t\rightarrow \infty} \frac{\EE[X(t)Y(t)]}{\rr_A^\circ \rr_B^\circ}= \frac{C_{xy}(0)}{\rr_A^\circ \rr_B^\circ},
\end{eqnarray*}
fluctuations will be relevant when $\Omega \sim N$, because then one can overcome the $N^{-1/2}$ dumping factor of the second-order corrections. The analysis of the possible values that $\Omega$ can take is difficult because the inner relationships that $\mu_{xx}$, $\mu_{xy}$, $\mu_{yx}$, $\sigma_x$, $\sigma_y$ and $\rho$ present. In fact the difficulty is inherited from $\gamma_A$, $\gamma_B$, $\alpha_{AA}$, $\alpha_{AB}$ and $\beta_{AB}$, which are neither bounded nor independent. Then, it is useful to introduce the following (final) re-parameterisation:
\begin{eqnarray*}
\alpha_{AA}&=&\frac{1}{ \chi}\frac{2}{\tau_0},\\
\alpha_{AB}&=& \frac{1}{\eta \chi}\frac{2}{\tau_0},\\
\beta_{AB}&=&\frac{\xi}{ \chi}\frac{1-\eta}{\eta} \frac{2}{\tau_0},\\
\gamma_A &=& \left[1+\frac{1-\chi}{\chi}\epsilon\right]\frac{2}{\tau_0},\\
\gamma_B&=& \xi\frac{1-\eta}{\eta} \frac{2}{\tau_0},
\end{eqnarray*}
where the four new variables $\chi$, $\epsilon$, $\eta$ and $\xi$ are in the $(0,1)$ range, and can be arbitrarily settled. With the proposed parameterisation all the constraints that affect the old parameters (the pure algebraic ones, as well as those coming from stability considerations) are identically satisfied,~\footnote{The new parameterisation proves incidentally that $\rho<1/2$.} and $\tau_0$ carries the characteristic time scale of the interactions at the microscopic level. 
The magnifying factors in the new parameters read
\begin{eqnarray*}
\Omega_{xx}&=& (1-\eta \epsilon)\frac{1-\chi}{\chi^2} + \frac{1}{2}\frac{1+\xi}{\xi} \frac{1}{\chi\eta}, \\
\Omega_{yy}&=&\left\{\left[\frac{1-\chi}{\chi}(1-\eta \epsilon)-1\right]\eta\xi+\frac{1+\xi}{2}\right\}\frac{1-\eta}{(1-\chi)\eta^2\epsilon}\\
&+&\frac{1}{2}\frac{1+\xi}{\xi} \frac{\chi}{(1-\chi)^2\eta\epsilon^2}, \\
\Omega_{xy}&=&-\frac{1}{2}\frac{1+\xi}{\xi}\frac{1}{(1-\chi)\eta\epsilon},
\end{eqnarray*}
and the stationary first-order solutions are $\rr_A^\circ = \chi$ and $\rr_B^\circ= (1-\chi)\eta \epsilon$. 

The first point to be noted is that no $\tau_0$  appears in any of the last expressions. So, the magnification effect does not depend on the characteristic time scale of the correlations. The second aspect of importance is that, for fixed values of $\chi$, $\epsilon$ and $\eta$, the magnifying factors become unboundedly large as $\xi \rightarrow 0$, and this parameter does not contribute to the value of $\rr_A^\circ$ and $\rr_B^\circ$. Then, magnification can be achieved for any (regular) value of the species stationary densities. Another favourable scenario is  $\rr_A^\circ\rightarrow 0$ and $\rr_B^\circ \rightarrow 0$: check for instance how for $\chi\rightarrow 0$, $\Omega_{xx}\rightarrow \infty$. This implies that the phenomenon is relevant in sparse systems as well, in spite of the fact that in such cases $N$ may eventually be very large. Note finally that magnification is not connected with the presence of oscillations of any peculiar frequency. On the one hand, the condition that determines that $T_0^{-1}$ replaces $\omega_0$ is 
\begin{equation*}
\xi<\frac{\chi \eta}{4(1-\chi)(1-\eta)\epsilon},
\end{equation*}
and as we have shown above $\xi\rightarrow 0$ always leads to magnification. This is reasonable since for a fixed $\tau_0$, $T_0$ increases the microscopic correlation range |see Appendix~\ref{AA}. On the other hand, for a fixed value of $\rr_A^\circ$, $\rr_B^\circ$ and $\xi$, $\omega_0$ embraces the whole positive real axis as $\eta$ varies. 
Therefore, in principle, one can consider models with either a large value for $\omega_0$, and reproduce the typical bid-ask bounce in a liquid market~\cite{MPMLMM05}, or a smaller one, and capture some seasonal character present in the market evolution, like in electricity markets~\cite{PMPSM06}.
Because, as it is shown in detail in Appendix~\ref{AA}, and it can be observed in Fig.~\ref{Spec_Fig}, the oscillatory behaviour is  also present in the second-order terms. 

\begin{figure}[htb]
{
\includegraphics[width=1.0\columnwidth,keepaspectratio=true]{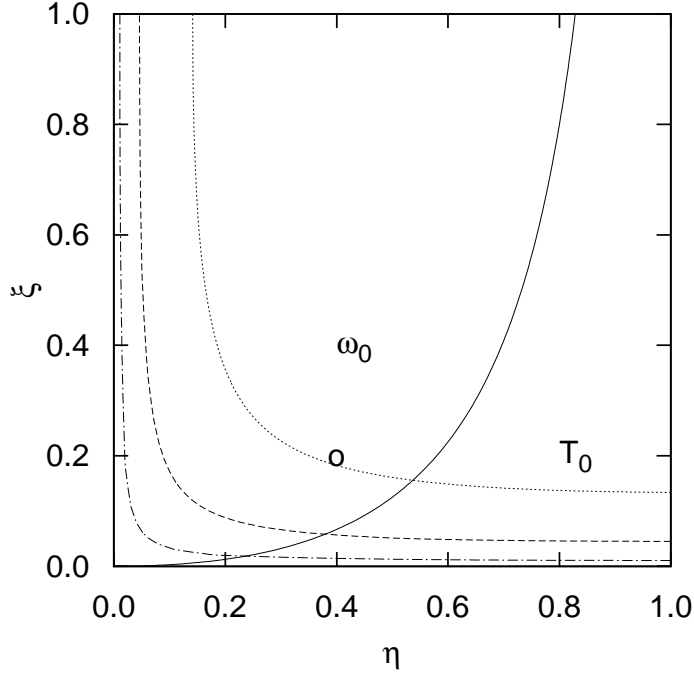}
}
\caption{Contour plot of the magnifying factor $\Omega_{zz}$. The magnifying factor values, to be compared with $N$, are 100 (dotted line), 250 (dashed line) and 1000 (dot-dashed line). The solid line is the borderline between the zones with oscillating behaviour ($\omega_0$) and without it ($T_0$).}
\label{Amp_Fig}
\end{figure}

Let us see magnification in a practical example. For clarity reasons we will condense the three magnifying factors defined above in a single plot. To this end we define $\Omega_{zz}$,
\begin{equation*}
\Omega_{zz}\equiv\Omega_{xx}+\Omega_{yy}-2\Omega_{xy}=\lim_{t\rightarrow \infty} \EE\left[\left(\frac{Y(t)}{\rr_B^\circ}-\frac{X(t)}{\rr_A^\circ}\right)^2\right],
\end{equation*}
a relevant quantity, as we will see below. Further, we will fix $\rr_A^\circ=\chi$, and change $\epsilon$ in a way that keeps $\rr_B^\circ$ constant, what leaves $\eta$ and $\xi$ as the only free parameters. In particular, we have set $\rr_A^\circ=\rr_B^\circ=0.2$, since we are interested in models in which the two sides are equally (no side is prioritised) and macroscopically populated. In Fig.~\ref{Amp_Fig} we observe some contour lines that represent configurations with the same amplification level, and how these lines cross the borderline that delimits those configurations with and without oscillating properties. Thus, for instance, we have marked with a small circle the location of the following parameter set: $\chi=0.2$, $\epsilon=0.625$, $\eta=0.4$, $\xi=0.2$. With this configuration $M=0.7 N$, and the amplification factor is about one hundred. 

\section{Price dynamics in a excess demand model}
\label{MM}
As we have stated above, the formula that will determine the price dynamics must depend on the nature of the species. Then it is time to identify $A$ and $B$ states and define how the evolution of the population of agents in each category translates into prices changes. Within this first model we will generically consider that $N_A(t)$ represents the offer and $N_B(t)$ represents the demand in a certain financial market, and assume that in such a market excess return reacts linearly to excess demand: a classical and ubiquitous point of view in the economic literature, {\it e.g.\/}~\cite{Z74,BG80,LM99,CB00,CCMZ01,D07}. Excess return measures the logarithmic earnings of the stock beyond the risk-free interest rate $r$, 
$R(t)\equiv\ln \left[S(t)e^{-r t}\right]$,
and excess demand is the difference between $N_B(t)$ and $N_A(t)$. Therefore we will have
\begin{eqnarray*}
\dd R(t)&=&\frac{\Xi}{N} (N_B-N_A) \dd t\\
&=&\Xi \left(\rr_B-\rr_A+\frac{Y-X}{\sqrt{N}}\right)\dd t \stackrel{t \gg \tto}{\longrightarrow}\Xi \left(\rr_B^\circ-\rr_A^\circ+\frac{Y-X}{\sqrt{N}}\right)\dd t,
\end{eqnarray*}
where $\Xi$ measures the sensitivity of prices to excess demand. This first pricing model is a good testing ground since the agent model will be responsible of any observed market property: we are simply integrating the differences in population.  

Let us consider the following paradigmatic example with $N=1000$ investors. We have set $\tau_0=10$ minutes, so it is of the same order of magnitude as a typical correlation length found in actual financial data~\cite{MMP00}. Beyond this, the rest of values were not based on actual market observations. In fact we have set $\chi=0.2$, $\eta=0.4$ and $\xi=0.2$, like in the example we emphasised in the previous Section, but slightly increased the value of $\epsilon$, $\epsilon=0.643$. This was intended to get $\rr_B^\circ \gtrsim \rr_A^\circ$, $\rr_B^\circ \approx 0.206$, whereas $\rr_A^\circ=0.2$. Note that $\rr_B^\circ-\rr_A^\circ>0$ characterises a growing economy in which wealth is injected into the market. This term is also responsible for a long-run exponential growth.

\begin{figure}[htb]
{
\includegraphics[width=0.75\columnwidth,keepaspectratio=true]{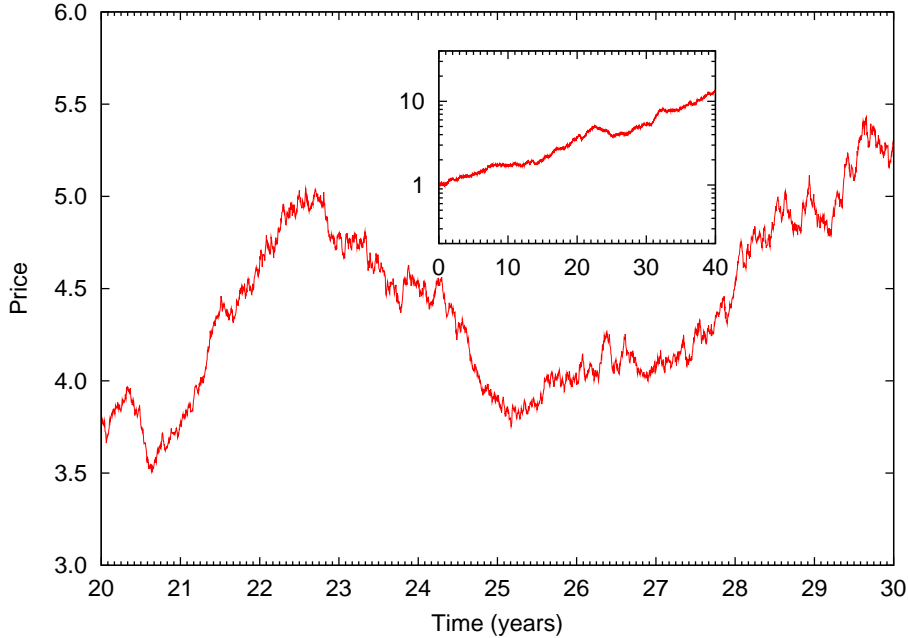}
}
\caption{(Color online) Time evolution of the daily closing value of (discounted) stock prices. We can see how the market suffers a market bubble (an upward trend followed by a downward trend) lasting 5 years and ending at the beginning of year 25. After that we find what is called a {\it sideways\/} trend, {\it i.e.\/} no trend at all, from the mid of year 25 to the mid of year 27, followed by a new upward trend with corrective movements in the middle. The inset shows the exponential growth in the long run.}
\label{Evol_Fig}
\end{figure}

A possible realisation of the dynamics of the species population was previously introduced in Fig.~\ref{Spec_Fig}, and in Fig.~\ref{Evol_Fig} we find the corresponding evolution of the stock price when $\Xi=10^{-3}$ min$^{-1}$. Here we sampled the complete data series to consider closing prices only, a usual practice in technical analysis. Moreover, in the confection of Fig.~\ref{Evol_Fig} and hereafter we are assuming that a trading day lasts 480 minutes, and that there are 250 trading days in a year. We observe in Fig.~\ref{Evol_Fig} the appearing of typical market charts: upward trends |an increasing succession of minima|, downward trends |a decreasing succession of maxima|, and sideways trends |a bouncing movement between two price levels. 

\begin{figure}[hbt]
{
\includegraphics[width=0.75\columnwidth,keepaspectratio=true]{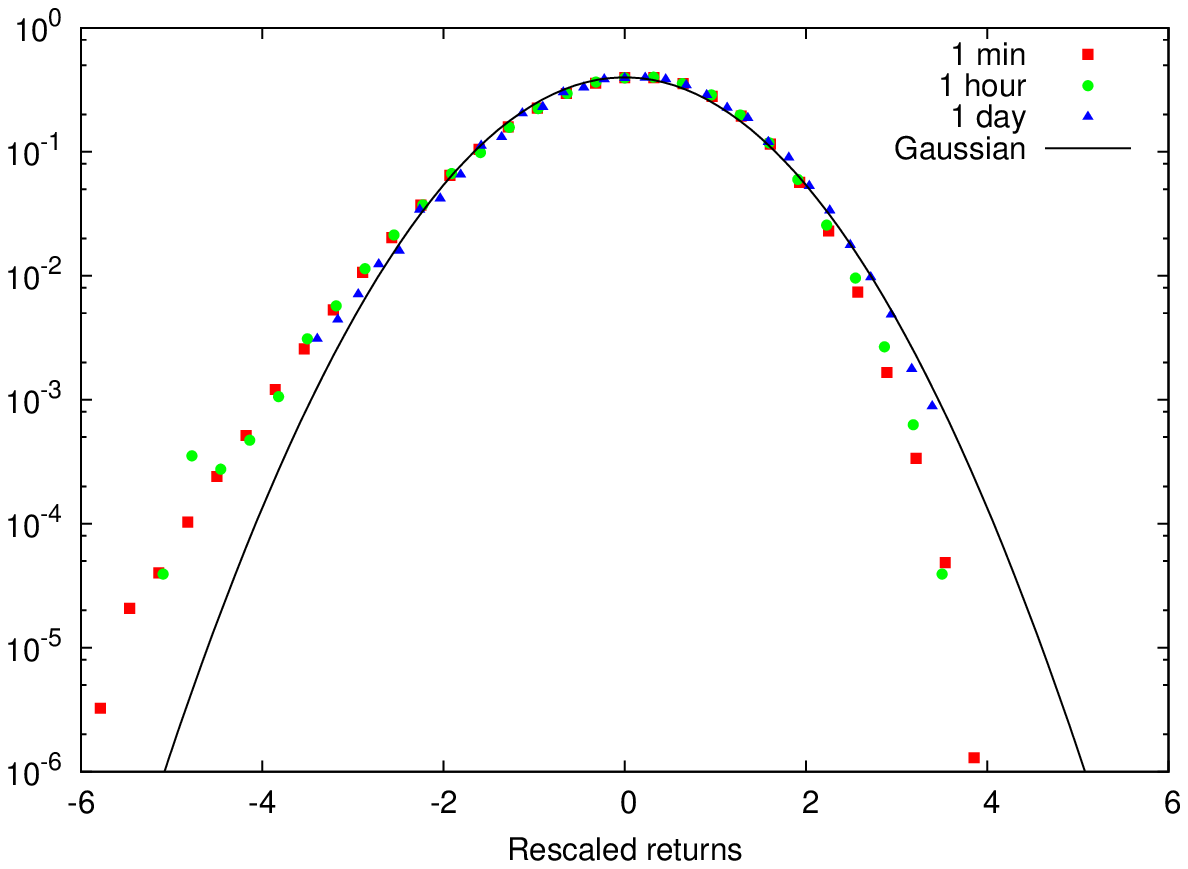}
}
\caption{(Color online) Fixed-horizon return behaviour. We can see how probability density functions at small time scales slightly but noticeably depart from Gaussian behaviour by exhibiting negative skew: the negative tail is fatter than the positive tail. Returns were divided by their sampling standard deviations to make them commensurable.}
\label{Pdf_Fig}
\end{figure}

\begin{figure}[htb]
{
\includegraphics[width=0.75\columnwidth,keepaspectratio=true]{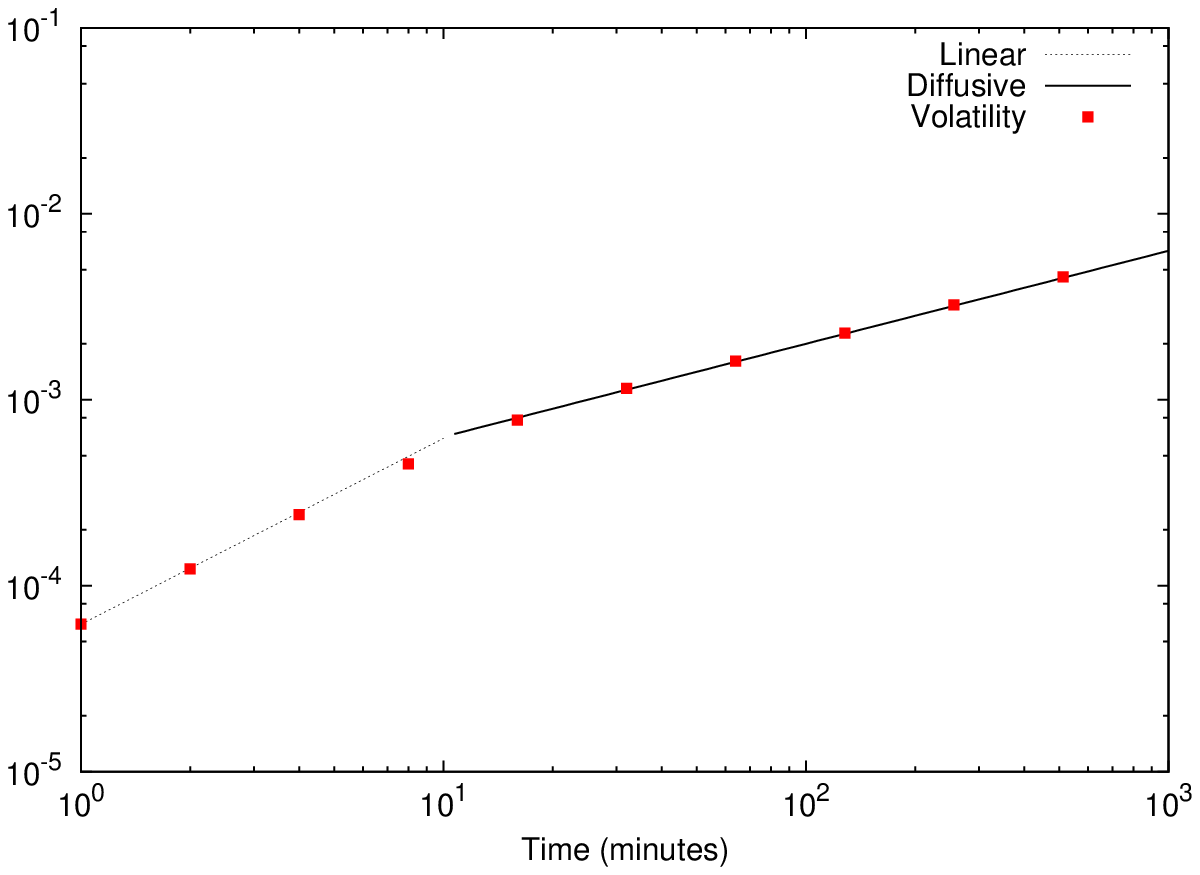}
}
\caption{(Color online) Volatility growth. In this figure we can see how the volatility growth presents two well different regimes. For $\tau<\tau_0$ the standard deviation of fixed-time returns shows (super-diffusive) linear growth whereas for $\tau>\tau_0$ it scales as $\sqrt{\tau}$, like in a diffusive process.}
\label{Volatility_Fig}
\end{figure}

In Fig.~\ref{Pdf_Fig} we present the outcome of a statistical analysis performed with the stationary data set of fixed-time returns  $R(\tau;t)=R(t+\tau)-R(t)$, $t>100$ minutes. We check that for $\tau \sim \tau_0$ correlations are important, Gaussian limit is not attained and skewness is observed, like in actual markets~\cite{MMP00,MS95,GPAMS99,C01}.    
This phenomenon is even more noticeable when  the standard deviation of fixed-time returns, a measure of the volatility of the market, is analysed, Fig.~\ref{Volatility_Fig}. Since $X(t)$ and $Y(t)$ are anti-correlated, and the return change is sensible to the difference of those magnitudes, we expect that volatility grows faster for small time scales, and reaches the diffusive regime for $\tau>\tau_0$. Traits of abnormal (both sub- and super-) diffusion have been reported to be present in real markets as well~\cite{MMP00,MMW03,MMPW06}.

In the confection of the previous plot we have used the complete set of returns available for each time scale $\tau$, by assuming the statistical equivalence of every sample $R(\tau;t)$ as a function of $t$. Moreover, the results above seem to indicate that, for $\tau \gg \tau_0$, the samples $R(\tau;t)$ and $R(\tau;t+\tau)$ ought to be also (almost) independent one from the other.  
\begin{figure}[htb]
{
\includegraphics[width=0.75\columnwidth,keepaspectratio=true]{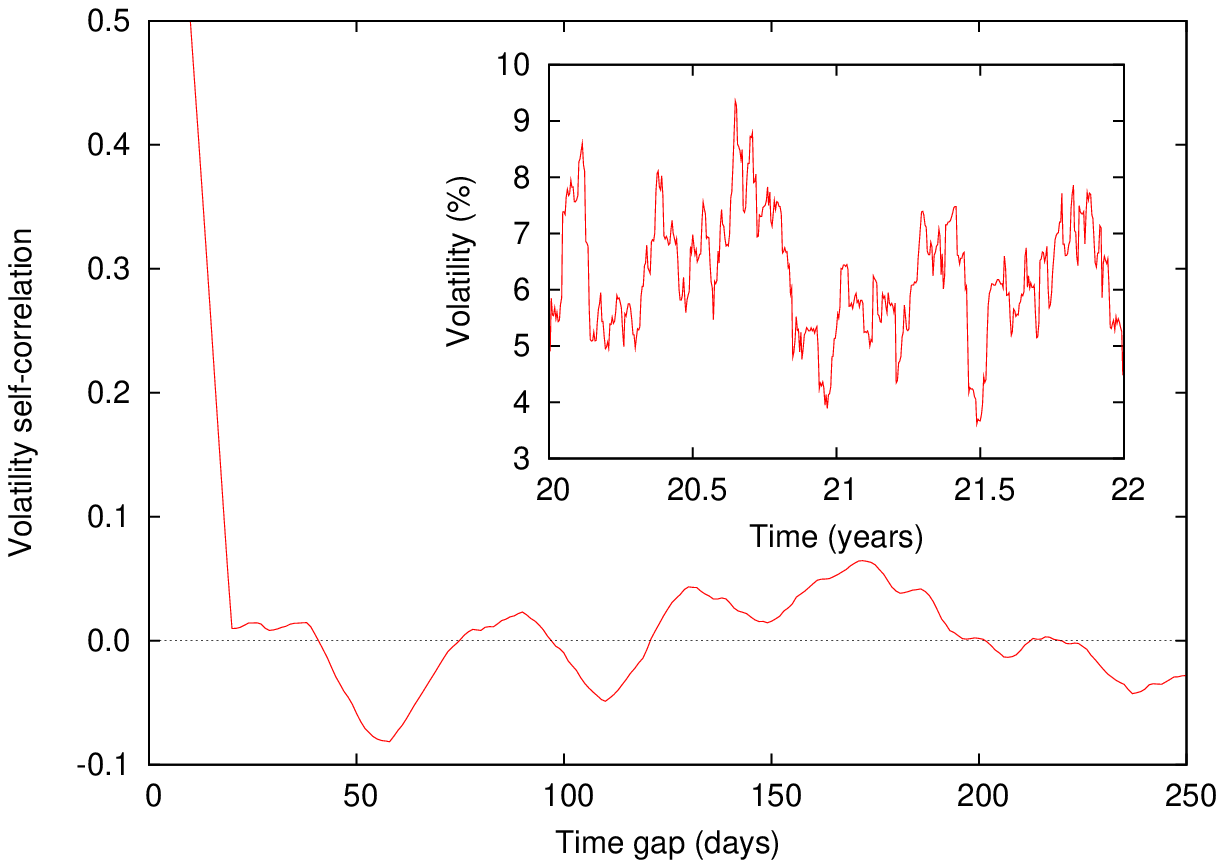}
}
\caption{(Color online) Volatility clustering. In this figure we can see how the one-month annualised volatility presents clustering (inset) and long memory. The increment of the correlation for times below 20 days is due to data overlapping.}
\label{Clustering_Fig}
\end{figure}
So, if we compute the realised $n$-$\tau$ volatility, $V_n(\tau;k)$:
\begin{equation*}
V_n(\tau;k)=\sqrt{\frac{1}{n}\sum_{m=1}^{n}\left[ R\Big(\tau;(k-m) \tau\Big)-\frac{1}{n} R\Big(n \tau;(k-n) \tau \Big)\right]^2},  
\end{equation*}
we should expect the outcome to be uniform in $k$, as well as an absence of correlation between $V_n(\tau;k)$ and $V_n(\tau;k+n)$. In order to check whether this assumption is true we have chosen $\tau=1$ day, and $n=20$ trading sessions, as a proxy for the one-month  realised volatility, a typical choice among practitioners.~\footnote{For instance, one-month volatilities are on the basis of VIX, the {\it Chicago Board Options Exchange\/} pioneering volatility index for the Standard \& Poor's 500 index.} The results were also {\it annualised\/}, which means here that they were increased by a factor $\sqrt{12.5}$, and only $k\geqslant 21$ are considered |we ignore the whole first day of simulation. The outcome, as it can be observed in the inset of Fig.~\ref{Clustering_Fig}, is that the market alternates long periods where the volatility is large, with periods of relative calm, a phenomenon known as {\it volatility clustering\/}~\cite{C01}. The presence clustering in the volatility is a well-documented feature of real markets that is usually explained in terms of the existence of volatility self-correlation. This correlation, as opposed to the return-to-return correlation, is long-ranged~\cite{L91} |confront time scales in Figs.~\ref{Clustering_Fig} and~\ref{Corr_Fig} below.     

\begin{figure}[h]
\includegraphics[width=0.75\columnwidth,keepaspectratio=true]{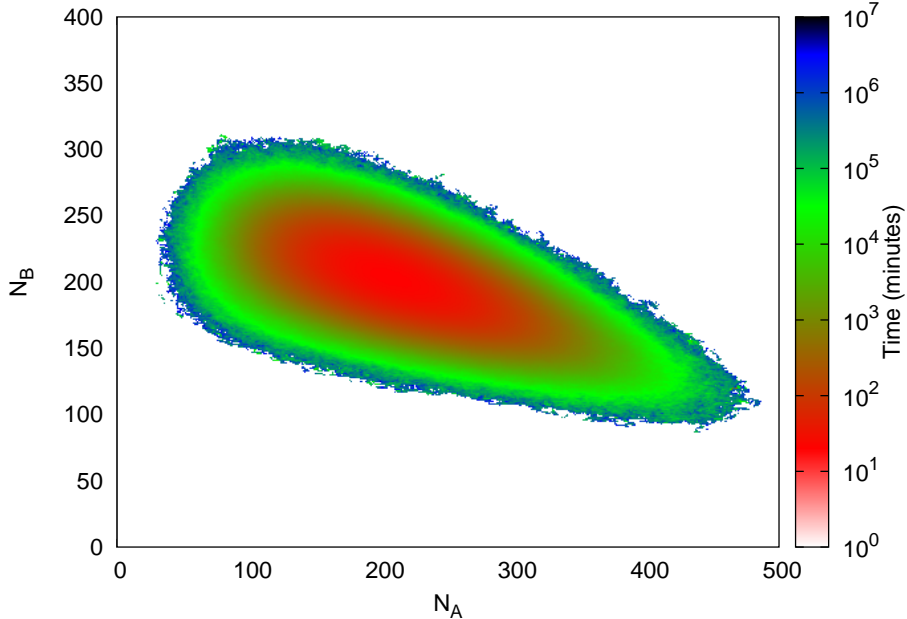}
\caption{(Color online) Phase diagram of the state of the system. We show the mean recurrence time, the mean time passed between two consecutive visits to the same state.
}
\label{Phase_Fig}
\end{figure}

In order to offer a plausible origin for this larger time scale, we have composed Fig.~\ref{Phase_Fig}. There we present, in a phase diagram, the mean recurrence time:  For each possible state of the system we have recorded all the visiting times, and performed a sample mean with the inter-event times. Therefore, at least two visits to a given state are needed in order to attach a non-zero value to that point. Once again we have disregarded the data within the first day. As it can be observed in this figure, the mean time grows in an exponential fashion when we depart from the stable fixed-point values, $N_A=200$ and $N_B=206$. Since the scale is logarithmic some lower bound must be chosen, and we have decided to remove those data points with a mean recurrence time smaller than 1 minute. This retains in the plot almost all non-zero values:~\footnote{Indeed, with this practice only few peripheral points were ignored: states that were visited several times in a rapid succession before the system leaved that zone and never returned to it.} the lowest recurrence time near the {\it core\/} is attained at $N_A=207$ and $N_B=203$, yielding a value of 20.17 minutes. As we can see in Fig.~\ref{Corr_Fig} again, this magnitude coincides with the time scale for which one-minute returns exhibit stronger anti-persistence. The slow decay in the volatility self-correlation may have thus its origins in those long periods needed by the system to return to the most outer zone, from where the largest absolute returns come: the light (green) ring marks a recurrence time of about 60 days, the time scale for which the volatility self-correlation is more intense |see Fig.~\ref{Clustering_Fig}. 

Therefore, the agent model is capable of conciliating short-range correlations in the microscopic level with long-range correlations in the macroscopic level, what may be linked with the presence of large business cycles in the financial data~\cite{BM46}.

\begin{figure}[htb]
{
\includegraphics[width=0.75\columnwidth,keepaspectratio=true]{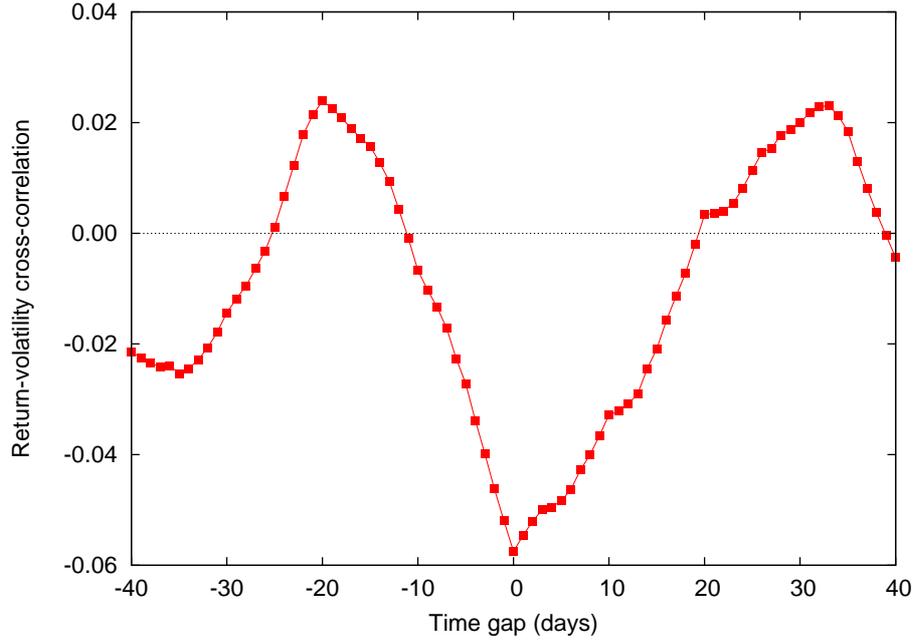}
}
\caption{(Color online) Leverage effect. In this figure we can see how 20-session returns and volatilities are negatively correlated, a phenomenon known as the leverage effect in the literature.}
\label{Leverage_Fig}
\end{figure}

Finally, another stylised fact that is commonly associated with clustering and long-range memory in the volatility is the so-called leverage or Fisher-Black effect~\cite{C01,C82,BMP01}. This phenomenon is generically characterised by a negative relationship between returns and volatilities. In our case, this effect can be barely observed when the cross-correlation between 20-day returns and volatilities are depicted |see Fig.~\ref{Leverage_Fig}. We must point out, moreover, that there are features shown in empirical studies related to this effect that are not detected in our example. For instance, from Fig.~\ref{Leverage_Fig} one cannot sustain the presence of a noticeable temporal asymmetry in the correlation, as expected~\cite{BMP01}. However, we can explain this departure from what is observed in actual markets on the basis of the usual interpretation of the leverage effect: the market digests with nervous the loses, and with confidence the rises. And we must remember at this point that the price information is not fed back into the species, so it is not possible such a reaction here. Therefore, the slight anti-correlation present in Fig.~\ref{Leverage_Fig} could be a side effect of the volatility self-correlation, 
but just some spurious result as well.

\section{Price dynamics in a liquidity model}
\label{Ext}
Another possible identification of the two species comes from considering that an agent in the $A$ state is a {\it liquidity provider\/}, whereas an agent in the $B$ state represents a {\it liquidity taker\/}~\cite{H07,BFL08}. Liquidity providers introduce limit orders into the market whereas liquidity takers operate through market orders.  Limit orders are orders with a limit price that represents the minimum (respectively maximum) price the investor is accepting for selling (respectively buying) a given number of shares, the volume of the order. The limit order is placed in the so-called limit order book, which is visible to rest of qualified investors and it remains there until one of the two following major events takes place: someone accepts the ask (respectively bid) price and the transaction is completed, or the investor removes the order from the book. 
Market orders are orders that automatically match the best opposite limit order out of the limit order book. This is just the way in which noise traders are assumed to participate in the market, but unlike them, liquidity takers survey the limit order book and look forward to a convenient trading opportunity, like predators do~\cite{VLMM07}. Note however that this results in transaction ($AB \rightarrow EE$), not in predation ($AB \rightarrow BB$). This and the rest of interactions determine how qualified agents migrate from one set to the other or remain inactive~\cite{HS96,HH07}.

This interpretation will be more relevant in liquid markets, where changes in offer or demand have a small impact on prices. In such a situation a relevant magnitude in the price formation mechanism is the spread, the difference between the lowest ask price and the higher bid price. Then, the pricing formula must connect the spread with the relative populations of liquidity providers and liquidity takers, but the issue is not so unambiguous as in the previous case.  
We have decided to use a pricing expression inspired by
~\cite{F02,FJ02}.
Consider these general guidelines:
\begin{enumerate}
\item The bigger the number of limit orders are in an order book, the lower the spread will be, and therefore, the lower the price change will be.
\item If the number of liquidity takers is small with respect to the number of liquidity providers, the price should tend to exhibit the typical bid-ask bounce pattern.
\item If the number of liquidity takers is large with respect to the number of liquidity providers, the most likely is that price shows a trend.  
\end{enumerate}

A feasible candidate that incorporates the above properties is the following discrete-time updating formula:
\begin{equation*}
R(t+\Delta t)=R(t)+\Xi\Theta\left(R(t)-R(t-\Delta t)\right)\left[\frac{N_B(t)}{N_A(t)}-\zeta\right]\Delta t, 
\end{equation*}
where $\Delta t$ is the time between two consecutive changes in the agents' state, $\Theta(\cdot)$ is the Heaviside step function and, for a matter of model simplicity, we will assume that $\zeta=\rr_B^\circ/\rr_A^\circ$. This liquidity model will share some traits with the previous excess demand model, since for large values of $N$ and $t$ we have
\begin{equation*}
R(t+\Delta t)\sim R(t)+\frac{\Xi}{\sqrt{N}}\Theta\left(R(t)-R(t-\Delta t)\right)\left[\frac{Y(t)}{\rr_B^\circ}-\frac{X(t)}{\rr_A^\circ}\right]\Delta t,
\end{equation*}
and the two formulas become very similar when $\rr_B^\circ \approx \rr_A^\circ$. The main distinguishing trait is the presence of the factor with the Heaviside step function, that may distort the evolution that the agent model dictates and introduce new properties.

\begin{figure}[htb]
{
\includegraphics[width=0.75\columnwidth,keepaspectratio=true]{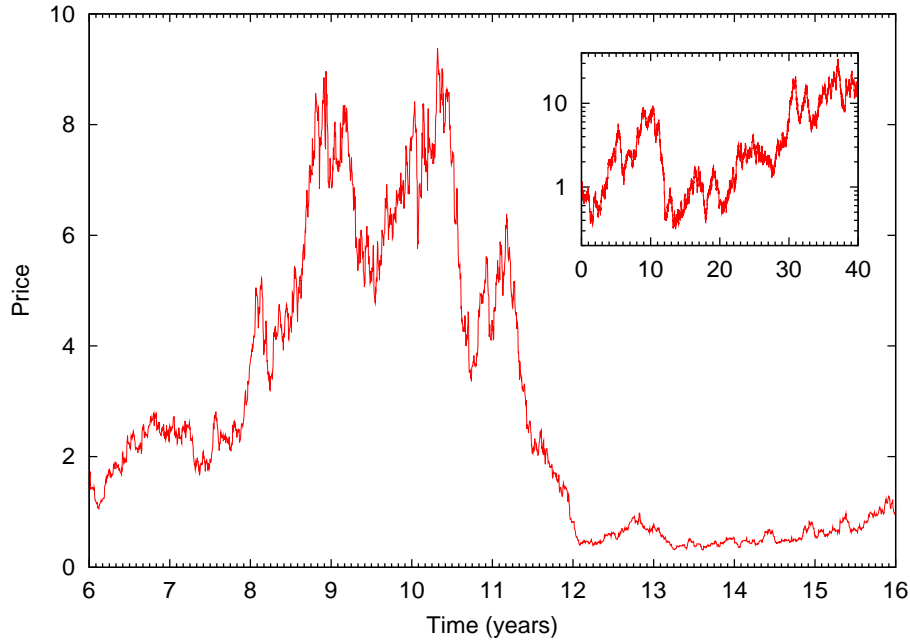}
}
\caption{(Color online) Time evolution of the daily closing value of (discounted) stock prices for the alternative dynamics based in the liquidity model. Among the several trends that are present, it is remarkable the sudden large drops which resemble market crashes.}
\label{Evol_liquid_Fig}
\end{figure}

In Fig.~\ref{Evol_liquid_Fig} we see how a sample time series mimics again a typical stock market evolution. In the confection of this plot we have kept the same parameters as in the previous market model with just one exception, the sensibility was set to $\Xi=0.05$ min$^{-1}$, with the aim of recovering a similar growth in the long run. However, we see how the market becomes much more volatile than it was in the previous case.

\begin{figure}[htb]
{
\includegraphics[width=0.75\columnwidth,keepaspectratio=true]{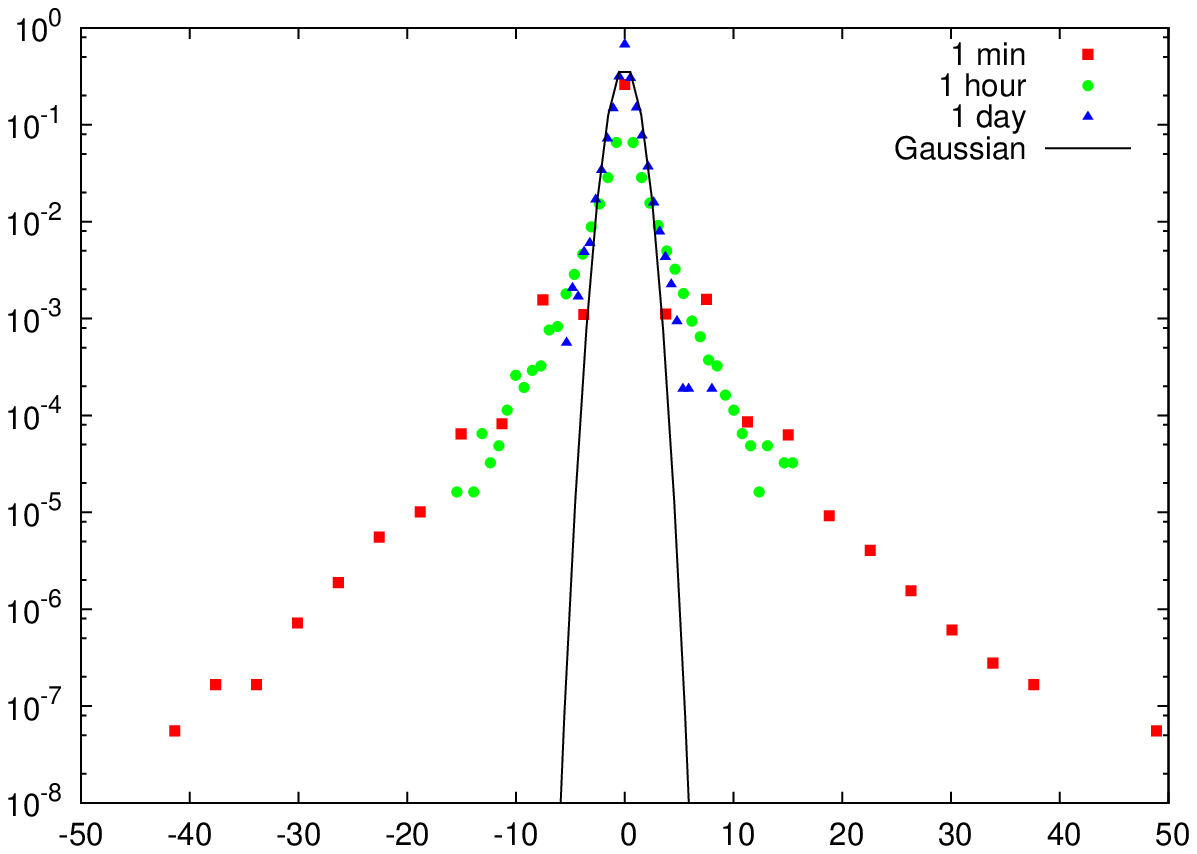}
}
\caption{(Color online) Fixed-horizon return behaviour. We can see how probability density functions present fat tails at intra-day time scales, and how the Gaussian behaviour is not fully recovered even in the case of daily returns.}
\label{Liquid_Pdf_Fig}
\end{figure}

\begin{figure}[htb]
{
\includegraphics[width=0.75\columnwidth,keepaspectratio=true]{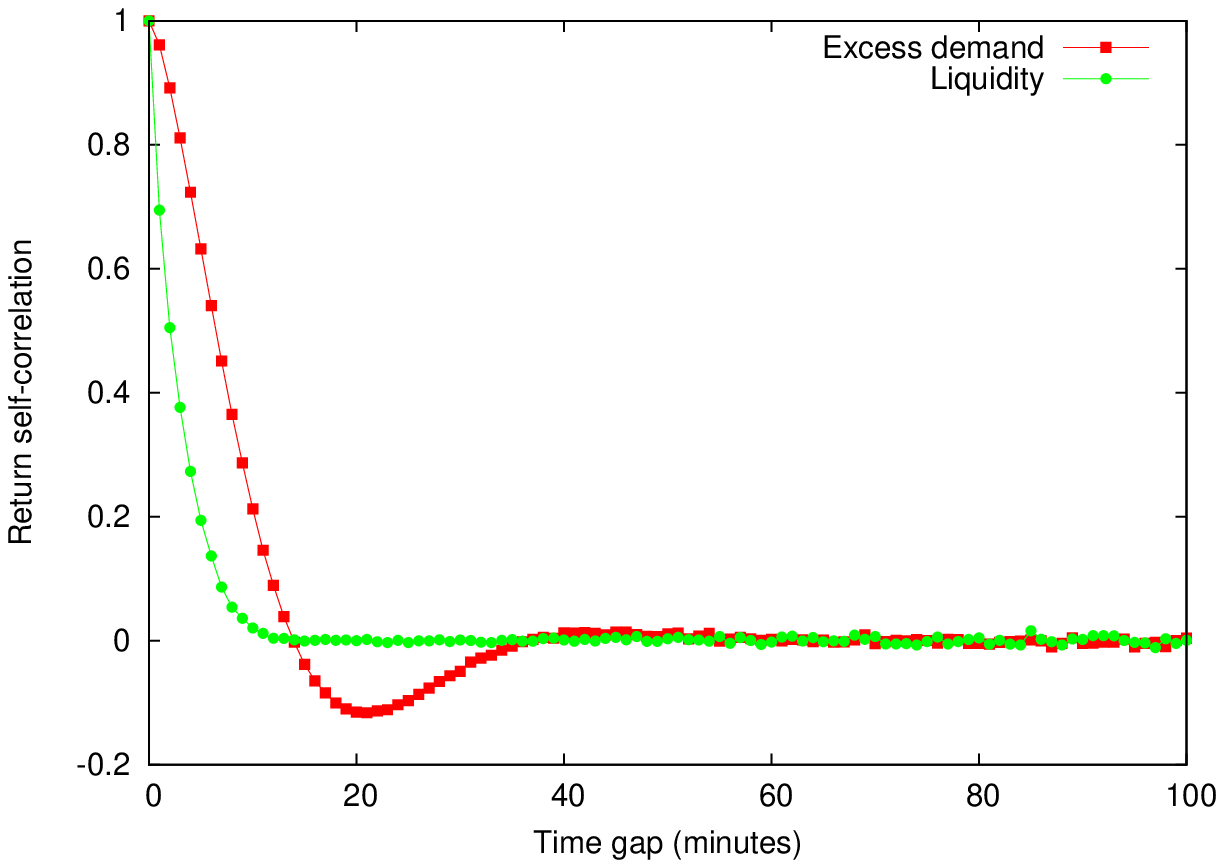}
}
\caption{(Color online) Linear correlation of one-minute returns for the two pricing models. The basic time scale $\tau_0=$ 10 minutes can be observed in both cases.}
\label{Corr_Fig}
\end{figure}

The impact of the new pricing dynamics in the probability density function of returns is also noticeable. In Fig.~\ref{Liquid_Pdf_Fig} one observes how one may find in practice changes amounting tens of standard deviations, like in actual markets~\cite{MMP00,MS95}. Finally, the correlation length is affected but time scale $\tau_0$ is preserved: in Fig.~\ref{Corr_Fig} we show a comparison between the one-minute-return correlation curves for both cases. On the contrary, the pricing mechanism in the liquidity model wipes completely the negative correlation that one can relate with $\omega_0$: the characteristic time scale of the oscillations is about 44.4 minutes with our parameter selection.

\section{Conclusions}
\label{Conclusions}
Along this article we have introduced a dynamical model that describes ultimately the evolution of financial prices. 
The main ingredient of the model is a finite set of identical interacting agents that at every moment can be accommodated into one of three excluding categories. The agents represent those traders whose operations may noticeably impact on the market, and the three available states characterise in a broad sense the investor possible attitude.

Active agents can spontaneously adopt a neutral attitude, but any other change is the outcome of an agent-to-agent interaction: agents may agree, or one agent can convince the other following hierarchical relationships. These simple rules encode a system in which the number of active agents may strongly fluctuate, thus overcoming the second-order nature of the effect. We have looked for the conditions that promote such amplification, and concluded that it does not depend on the time scale of the interactions, can be obtained for any choice of the first-order stationary densities |even though it is more relevant in sparse systems|, and it is not the result of a resonance.

Once we have analysed the dynamics of the agent instantaneous properties, we have moved into the pricing problem. We have considered two different identification of the categories and, in each case, a suitable pricing expression is set. We have simulated the time evolution of the asset price for representative values of the involved parameters. We have shown how sample realisations reproduce several stylised facts reported in actual financial data sets: the price evolution displays upward, downward and sideways trends; probability density functions of small time-scale returns present fat tails and skewness; volatility behaves accordingly in a non-diffusive way within the same time horizon, and presents clustering in a larger time scale; and traces of some leverage effect can be found.

In a forthcoming work we are planning to explore how the properties shown by the agent model depend on the assumptions made, to refine its connections with actual financial systems, and to consider further alternative interpretations that may be relevant in market dynamics not considered here.

\acknowledgments
The author acknowledges partial support from the former Spanish {\it Ministerio de Educaci\'on y Ciencia\/} under contract no. FIS2006-05204-E, from the {\it Generalitat de Catalunya\/} under contract no. 2005 SGR-00515, and from the {\it Junta de Castilla y Le\'on\/} under contract no. SA034A08.
\appendix
\section{Correlation functions}
\label{AA}
The cross-correlation theorem states that, for any couple of two random variables $X(t)$ and $Y(t)$ we can compute its stationary auto-correlation function though:
\begin{equation*}
C_{xy}(\tau)\equiv\lim_{t\rightarrow \infty} \EE[X(t)Y(t+\tau)] =\int_{-\infty}^{\infty} \frac{\dd \omega}{2 \pi} P_{xy}(\omega) e^{-i \omega \tau},
\end{equation*}
where
\begin{equation*}
P_{xy}(\omega)=\lim_{t\rightarrow \infty}\int_{-\infty}^{\infty} \frac{\dd \omega'}{2 \pi} \EE[\tilde{X}^*(\omega)\tilde{Y}(\omega')]  e^{-i (\omega'-\omega) t},
\end{equation*}
$\tilde{X}(\omega)$ stands for the Fourier transform of $X(t)$, and so forth. When $Y(t)$ coincides with $X(t)$, $P_{xx}(\omega)$ is termed the power spectral density function of $X(t)$, and cross-correlation theorem is known as the Wiener-Khinchin theorem. 

In our case equations~(\ref{dX}) and~(\ref{dY}) lead to
\begin{equation*}
P_{xx}(\omega)=\frac{\mu^2_{xy} \sigma^2_y+\sigma^2_x \omega^2}{\omega^2 \mu^2_{xx}+ \left(\omega^2 -  \mu_{xy}\mu_{yx}\right)^2},
\end{equation*}
\begin{equation*}
P_{yy}(\omega)=\frac{\mu^2_{yx} \sigma^2_x+\mu^2_{xx} \sigma^2_y-2\rho \mu_{xx}\mu_{yx} \sigma_x\sigma_y +\sigma^2_y \omega^2}{\omega^2 \mu^2_{xx}+ \left(\omega^2 -  \mu_{xy}\mu_{yx}\right)^2},
\end{equation*}
\begin{eqnarray*}
P_{xy}(\omega)&=&\frac{-\mu_{xx}\mu_{xy} \sigma^2_y+\mu_{xy}\mu_{yx}\rho \sigma_x \sigma_y}{\omega^2 \mu^2_{xx}+ \left(\omega^2 -  \mu_{xy}\mu_{yx}\right)^2}\\
&+&\frac{i \left(\mu_{yx} \sigma^2_x+\mu_{xy} \sigma^2_y-\mu_{xx}\rho \sigma_x \sigma_y\right)\omega-\rho \sigma_x \sigma_y \omega^2}{\omega^2 \mu^2_{xx}+ \left(\omega^2 -  \mu_{xy}\mu_{yx}\right)^2}.
\end{eqnarray*}
Note that every function has the same basic structure, namely
\begin{equation*}
P(\omega)=\frac{\kappa_1+i \kappa_2 \omega + \kappa_3 \omega^2}{\omega^2 \mu^2_{xx}+ \left(\omega^2 -  \mu_{xy}\mu_{yx}\right)^2},
\end{equation*}
therefore we can compute the auto- and cross-correlation functions at once,
\begin{eqnarray*}
C(\tau)&=&\Bigg[\frac{\kappa_1+\kappa_3 \mu_{xy}\mu_{yx}}{\mu_{xx} \mu_{xy}\mu_{yx}}\cos(\omega_0 \tau)+ \frac{\kappa_1-\kappa_3 \mu_{xy}\mu_{yx}}{2 \mu_{xy}\mu_{yx}\omega_0}\sin(\omega_0 |\tau|)\\
&+&\frac{\kappa_2}{\mu_{xx} \omega_0} \sin(\omega_0 \tau)\Bigg]\frac{e^{-|\tau|/\tau_0}}{2},
\end{eqnarray*}
where $\omega_0 \in  \RR^{+}$ coincides with that defined in~(\ref{omega0}), and $\tau_0$ was introduced in (\ref{tau_0}). If $T_0 \in  \RR^{+}$, see Eq.~(\ref{T0}), we have instead
\begin{eqnarray*}
C(\tau)&=&\Bigg[\frac{\kappa_1+\kappa_3 \mu_{xy}\mu_{yx}}{\mu_{xx} \mu_{xy}\mu_{yx}}\cosh(\tau/T_0)+ \frac{\kappa_1-\kappa_3 \mu_{xy}\mu_{yx}}{2 \mu_{xy}\mu_{yx}}T_0\sinh(|\tau|/T_0)\\
&+&\frac{\kappa_2}{\mu_{xx}} T_0\sinh(\tau/T_0)\Bigg]\frac{e^{-|\tau|/\tau_0}}{2},
\end{eqnarray*}
and two correlation time scales appear. Finally note that the corresponding variances and co-variances are obtained after setting $\tau=0$.


\end{document}